\documentclass{article}

\usepackage[preprint]{neurips_2026}
\workshoptitle{SaTQuML: Secure and Trustworthy Quantum Machine Learning}

\usepackage[utf8]{inputenc}
\usepackage[T1]{fontenc}
\usepackage{amsfonts}
\usepackage{amsmath}
\usepackage{amssymb}
\usepackage{hyperref}
\usepackage{cleveref}        
\usepackage{url}
\usepackage{natbib}          
\usepackage{braket}          
\usepackage{booktabs}
\usepackage{nicefrac}
\usepackage{microtype}
\usepackage{xcolor}
\usepackage{graphicx}
\usepackage{subcaption}
\usepackage{multirow}
\usepackage{array}
\usepackage{tikz}
\usetikzlibrary{positioning,arrows.meta,fit,backgrounds,shapes.geometric,calc}

\newcommand{\Itrue}{I_{\rm true}}
\newcommand{\Irep}{I_{\rm reported}}
\newcommand{\Bphys}{B_{\rm physical}}

\newcommand{\rhonv}{\rho_{\rm NV}}
\newcommand{\SU}{\mathrm{SU}}

\title{When Measurement Constraints Favor Quantum Computational Sensing for Stealthy Power-Grid Attack Detection}

\author{%
  \textbf{Saisubramaniam Gopalakrishnan},
  \textbf{Supreeth B S},
  \textbf{Pranav Sarda}, \\
  \textbf{Ashesh Xalxo},
  \textbf{Dagnachew Birru} \\
  Phi Labs, Quantiphi \\
  \texttt{gopalakrishnan.saisubramaniam@quantiphi.com}
}

\begin{document}

\maketitle

\begin{abstract}
Power-grid defenses that rely on digital telemetry remain vulnerable to
stealthy attacks that preserve plausible reported states while altering the
underlying physical system. We study when Nitrogen-Vacancy (NV) sensing provides a useful independent
physical channel, and when coherent processing before measurement adds value.
Across IEEE 14-, 30-, and 118-bus simulations with Lindblad NV models, we
evaluate standard FDIA, BDD-stealth, statistical-stealth, and concealed
topology attacks. The results reveal an observability hierarchy: evidence
shifts from digital telemetry, to reported-versus-physical consistency, to the
physical NV state.  Quantum Computational Sensing (QCS)  follows this selectivity, becoming informative only
when the physical state itself carries attack evidence.
We then compare QCS, conventional 4-setting NV
readout, and tomography under matched measurement budgets. At a matched total
budget of only 40 physical trials per sensor on case14, QCS reaches
AP $0.944$, versus $0.800$ for conventional NV readout and $0.751$ for
tomography; the same low-budget ordering holds on case30 and case118.
Multi-setting methods recover as additional measurements become affordable,
showing that the QCS benefit is a measurement-efficiency advantage rather
than a universal accuracy advantage.
Finally, we ask where the benefit varies across the three case simulations. Although the
quantum-to-classical Fisher-information ratio increases from $1.21\times$ to
$1.54\times$, Normal--Attack Helstrom separation collapses in the harder
regimes, and interleaved control substantially increases that separation only
on case14, thereby quantum sensitivity does not necessarily imply
task-relevant distinguishability. Our results show that realized QCS utility
depends on the full chain from physical perturbation to state separation,
coherent processing, and measurement under the resource constraints of the
task.
\end{abstract}

\section{Introduction} 
\label{sec:introduction} 
 

Modern power-grid monitoring relies on digitally reported  Supervisory Control and Data Acquisition (SCADA) telemetry for
state estimation and downstream control. Under false-data injection, however,
an adversary can manipulate those measurements while preserving the internal
consistency expected by conventional bad-data detectors. In the canonical
linearized setting, attacks
can remain unobservable to residual-based
detection \citep{liu2011false}. Data-driven attackers can similarly construct
stealthy directions from observed measurement traces without requiring complete
knowledge of the grid model
\citep{lakshminarayana2021datadriven,tian2022datadriven}.
These attacks expose a basic limitation: \emph{digital plausibility is not
physical correctness}. If the observation channel itself is manipulated,
a stronger downstream classifier cannot necessarily recover the hidden physical
information.
 
A natural response is to corroborate cyber-reported measurements with an
independent observation generated directly by the physical system, an idea
with precedent in smart-grid physical attestation
\citep{roth2013physicalattestation}. 
Quantum sensing has already been explored as part of power-grid cybersecurity,
including architectures that combine quantum sensing, secure communication,
and downstream machine learning \citep{nader2026triplelayer}. This motivates
quantum sensing as an independent physical-security channel, but leaves a
different QML question open: \emph{once the sensor state itself is quantum,
is there value in processing that state coherently before measurement?}
A quantum sensor can improve security simply
because it supplies an additional physical observation; that benefit does not
by itself establish a quantum-processing advantage. We therefore separate
\emph{the value of independent physical sensing} from
\emph{the value of coherent processing of the quantum sensor state}.

Quantum computational sensing (QCS) \citep{khan2026qcs} provides a natural
framework for studying this second question. Rather than first converting a
quantum sensor state into a classical representation and then classifying it,
QCS interleaves sensing with trainable coherent processing before a final
measurement. Given labeled normal and attack examples, the coherent-control
parameters are optimized so that the final measurement directly separates the
classes. The trainable control is therefore part of the sensing process itself:
QCS learns directly from the native quantum state generated by the physical
sensor and shapes it toward a task-specific final measurement. This makes the
setting quantum-native QML: the model operates on a physically generated
quantum state, rather than on classical features that must first be encoded
into a quantum circuit. This is particularly relevant
when physical measurement resources are limited: conventional readout and
tomography spend repeated state preparations across several measurement
settings, whereas QCS learns a task-specific final measurement.
\begin{figure}[t]
\centering
\includegraphics[width=\linewidth]{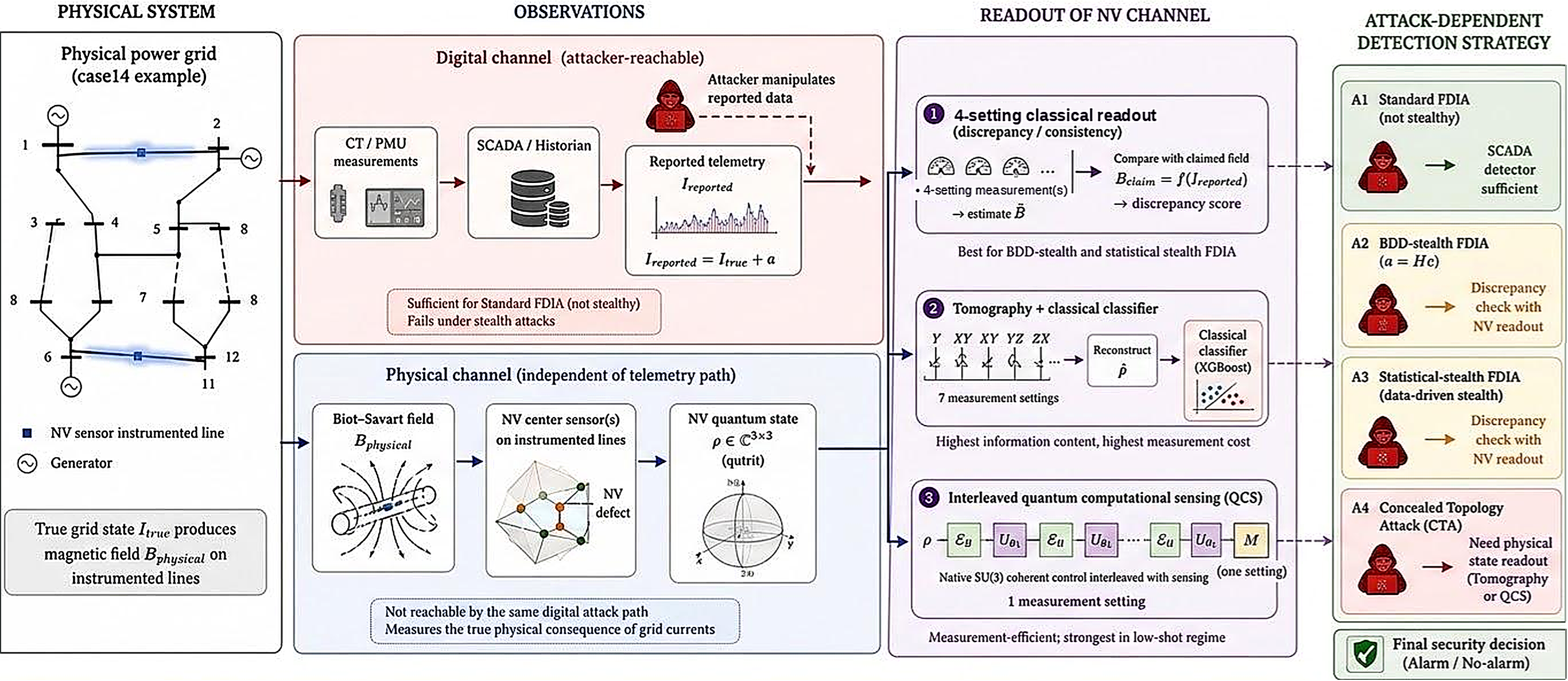}
\caption{The grid produces digital SCADA
telemetry, which can be manipulated by stealth attacks, and an independently
generated NV sensing channel. Different attacks leave different evidence
available. When the physical NV state carries the relevant signal, we compare
conventional NV readout, tomography, and interleaved QCS under matched sensing
resources.}
\vspace{-0.3cm}
\label{fig:overview}
\end{figure}

In this work, we consider Nitrogen-Vacancy (NV) quantum sensors positioned
near selected transmission lines, where the true line current generates a
magnetic field that directly modulates the NV spin state. Because this physical
signal is generated independently of the reported SCADA value, it can provide
evidence unavailable from compromised telemetry alone.

\Cref{fig:overview} summarizes the framework. Our evaluation is organized
around three questions:

\textbf{RQ1 -- Attack-dependent observability:}
Which evidence channel remains informative under different attack mechanisms?
We compare digital telemetry, reported-versus-physical consistency, and the
physical NV state across standard false-data injection, model-aware stealth,
statistical-stealth, and concealed topology attacks. The results reveal a
channel-dependent hierarchy: digital telemetry exposes standard FDIA,
reported-versus-physical consistency exposes BDD-stealth, while concealed
topology attacks shift the useful evidence to the physical sensor state.

\textbf{RQ2 -- Measurement-constrained quantum utility:}
When the physical NV state carries the relevant signal, how should a limited
measurement budget be spent? We compare QCS, conventional NV readout, and
tomography under both matched total physical trials and matched per-setting
precision. QCS is most competitive when the total measurement budget is
constrained, where its single-setting readout avoids splitting measurements
across multiple bases; richer multi-setting readout recovers as additional
trials become affordable.

\textbf{RQ3 -- Quantum sensitivity versus task-relevant distinguishability:}
What limits the conversion of quantum sensing information into detection
performance? Across IEEE 14-, 30-, and 118-bus systems, the
quantum-to-classical Fisher-information ratio increases from $1.21\times$ to
$1.54\times$, yet the Normal--CTA Helstrom separation falls sharply in the
harder sensing regimes. Thus, greater quantum sensitivity does not necessarily
imply greater class separability. The combined Fisher-information, Helstrom,
readout, and control analyses identify physical state separation and its
accessibility to the learned measurement as distinct limits on realized QCS
utility.
\section{Related Work}
\label{sec:related-work}

\subsection{Stealth attacks and physical attestation}
\label{sec:rw-attestation}

False-data injection attacks (FDIA) can manipulate grid telemetry while
preserving consistency with the defender's digital model. Prior work studies
both model-aware stealth attacks designed to evade residual-based bad-data
detectors \citep{liu2011false} and data-driven variants that infer stealthy
perturbations directly from measurement traces
\citep{lakshminarayana2021datadriven,tian2022datadriven}. Concealed topology
attacks extend this threat by masking real physical network changes behind
falsified telemetry
\citep{onebreaker2015,localcyberphysical2018,topologylearning2021}.
Physical attestation addresses this limitation by introducing an independent
physical observation against which cyber-reported state can be checked
\citep{roth2013physicalattestation}. Our work builds on this principle but asks
a different question: when that independent observation is generated by a
quantum sensor, what benefit comes from the physical sensing channel itself,
and what additional benefit, if any, comes from coherently processing the
sensor state before measurement?

\subsection{QML and quantum sensing for grid security}
\label{sec:rw-qml-grid}

Recent QML approaches to grid cybersecurity apply variational quantum
classifiers or quantum kernels to classically reported SCADA features for
intrusion and FDIA detection
\citep{ogiesoba2025qml,thomos2026vqc,cultice2025qsvm}. These works investigate
quantum models for downstream classification, but their inputs remain derived
from the same digital telemetry whose integrity is under attack.
Closer to our setting, \citet{nader2026triplelayer} combine quantum-secured
communication, distributed quantum sensing, and downstream machine learning
for power-grid AGC security. Their work establishes an important precedent for
using quantum sensing as an independent physical-security channel in the grid.
The distinction in our work is where learning occurs: rather than measuring the
quantum sensor first and applying learning downstream, we ask whether the
sensor's quantum state should itself undergo task-specific coherent processing
before readout. This separates the security value of \emph{having} a quantum
sensor from the QML value of \emph{processing its native quantum state}.

\subsection{Quantum computational sensing and distributed quantum sensing}
\label{sec:rw-qcs}

Quantum computational sensing (QCS) \citep{khan2026qcs} provides the
methodological basis for this question. QCS interleaves physical sensing
evolution with trainable coherent control so that the final measurement is
optimized directly for a downstream task, rather than first reconstructing a
general-purpose classical representation of the sensor state. We instantiate
this paradigm on the native spin-1 state of an NV center and evaluate it under
an adversarial cyber-physical sensing task with explicit physical measurement
constraints.

A complementary literature studies distributed and entangled quantum sensor
networks. Supervised-learning-assisted entangled sensing and related
experimental work show that coherent processing across sensors can improve
classification performance
\citep{zhuang2019slaen,xia2020variational,bosonic2024controllable}. Such
architectures assume access to shared quantum resources across sensing nodes.
Our primary deployment model instead processes spatially separated NV sensors
locally and combines only their classical outputs, avoiding an inter-sensor
quantum-networking requirement; joint coherent processing is evaluated only as
an extended ablation.
Native multilevel control of NV centers and NV magnetometry are themselves
established capabilities
\citep{nvqutritgates2026,multilevelsensing2025,hoegen2026review}. We use these
capabilities to keep the sensing and computational Hilbert spaces physically
aligned. The resulting question is therefore not whether quantum sensing or
qutrit control is possible, but \emph{when task-specific processing of a
naturally generated quantum sensor state is preferable to conventional
readout under the same physical measurement budget}.

\section{System Model and Quantum Framework}
\label{sec:system-framework}

\subsection{Threat and System Model}
\label{sec:threat-model}

\textbf{Observation channels and edge-trust boundary:} Let $\Itrue$ denote the true line-current state. The defender observes the
system through two parallel channels. Conventional SCADA telemetry provides a
digitally accessible vector $\Irep$ that may be manipulated by an attacker.
Independently, instrumented transmission lines generate a physical sensing path $
    \Itrue \rightarrow \Bphys \rightarrow
    \rhonv\in\mathbb{C}^{3\times3},
$
where $\Bphys$ is the magnetic field produced by the actual current and
$\rhonv$ is the resulting NV spin-1 state. Neither $\Bphys$ nor $\rhonv$ is
directly available as a classical variable; information must be extracted
through physical measurement of the sensor state.

The NV center, its local coherent controls, and photon-count readout are
assumed to reside within a trusted sensing appliance. Quantum processing
therefore acts locally on the native sensor state before measurement, without
requiring transfer to a remote quantum processor or a distributed quantum
network. Once a classical outcome leaves this boundary, authenticated
classical transport is assumed.

\textbf{Attack taxonomy:}
We consider four threat models chosen to stress distinct evidence channels:

\begin{enumerate}
    \item \textbf{Standard FDIA:}
    Direct perturbation of the reported measurements,
    $\Irep'=\Irep+a$, without an explicit stealth objective
    \citep{liu2011false}.

    \item \textbf{Model-aware BDD-stealth FDIA:}
    Structured injections $
        a = Hc $
    chosen so that the attacked measurements remain consistent with a shifted
    state estimate and preserve the residual used by Bad-Data Detection (BDD)
    \citep{liu2011false}.

    \item \textbf{Statistical-stealth FDIA:}
    Data-driven perturbations constructed from the normal-measurement
    distribution without requiring complete knowledge of the grid model
    \citep{lakshminarayana2021datadriven,tian2022datadriven}.

    \item \textbf{Concealed topology attack (CTA):}
    A physical topology change modifies the true system while falsified
    telemetry preserves an apparently normal digital view:
    $
        \mathcal{G}_{\mathrm{true}}\neq\mathcal{G}_{\mathrm{reported}}.
    $
    We additionally suppress the naive reported-versus-physical discrepancy,
    leaving the physical sensor state as the only informative channel
    \citep{onebreaker2015,localcyberphysical2018,topologylearning2021}.
\end{enumerate}

\begin{table}[t]
\centering
\caption{Attack-dependent evidence availability across digital and physical
sensing channels.}
\label{tab:observability-matrix}
\resizebox{\linewidth}{!}{%
\begin{tabular}{lccc}
\toprule
Attack & Digital telemetry &
Reported-vs-physical consistency & Physical NV state \\
\midrule
Standard FDIA             & informative & informative & not required \\
Model-aware stealth FDIA  & weak        & informative & not required \\
Statistical-stealth FDIA  & weak        & informative & not required \\
Concealed topology attack & weak        & suppressed  & informative \\
\bottomrule
\end{tabular}}
\end{table}

Here, $\Irep$ denotes reported telemetry, $a$ the injected attack vector,
$H$ the state-estimation measurement Jacobian, $c$ the attacker-chosen state
perturbation, and $\mathcal{G}$ the grid topology. \Cref{tab:observability-matrix} summarizes the hypothesis underlying our
evaluation. For cyber-only FDIAs, the physical current is unchanged, so the
physical channel is useful primarily for exposing inconsistency between the
reported and measured worlds. CTA instead makes the physical NV state itself
the relevant evidence source for comparing alternative quantum-sensing
strategies.

\subsection{Quantum Sensing and Computational Framework}
\label{sec:architecture}

\textbf{Edge-native NV sensing:}
The NV electronic ground state is a native spin-1 qutrit,
$\rho\in\mathbb{C}^{3\times3}$. We model its field-dependent evolution with a
Lindblad master equation incorporating zero-field splitting, Zeeman coupling,
and relaxation/dephasing. Because the sensing and computational
Hilbert spaces coincide, coherent processing acts directly on the physical NV
state rather than on classical features that are subsequently re-encoded into
a separate quantum circuit.

We consider three routes from the same physical NV sensing process to a
security decision. Here, $\rho_B$ denotes the NV state generated by the local
magnetic field, $M_{1:m}$ a set of $m$ measurement settings, $f_\phi$ a
downstream classical classifier, and $\hat y$ the predicted label.

\begin{enumerate}
    \item \textbf{Conventional NV readout.}
    Finite-shot measurements produce classical field-sensitive features,
    followed by downstream classification:
    \[
        \rho_B
        \xrightarrow{\;M_{1:m_{\rm readout}}\;}
        x_{\rm NV}
        \xrightarrow{\;f_\phi\;}
        \hat y .
    \]

    \item \textbf{Tomographic readout.}
    Multi-basis measurements first reconstruct the qutrit state and then
    classify it classically:
    \[
        \rho_B
        \xrightarrow{\;M_{1:m_{\rm tomo}}\;}
        \hat\rho
        \xrightarrow{\;f_\phi\;}
        \hat y .
    \]

    \item \textbf{Quantum computational sensing.}
    Following the QCS architecture of \citet{khan2026qcs}, field-dependent
    sensing is interleaved with trainable coherent control before a final
    measurement. For our open-system NV sensor,
    \[
        \rho_{k+1}
        =
        U_{\theta_k}
        \mathcal{E}_{B,\Delta t}(\rho_k)
        U_{\theta_k}^{\dagger},
    \]
    where $\mathcal{E}_{B,\Delta t}$ is the Lindblad sensing channel over
    $\Delta t$ and $U_{\theta_k}\in\mathrm{SU}(3)$ is the trainable qutrit
    control. After $N$ sensing/control stages, the state is measured once.

    For spatially separated sensors, our primary architecture processes each
    qutrit locally,
    \[
        \rho_i \rightarrow \mathrm{QCS}_i \rightarrow z_i ,
        \qquad
        [z_1,\ldots,z_K] \rightarrow \hat y. 
    \]
    We additionally evaluate a \emph{trainable Positive-Operator-Valued Measure
(POVM) readout}, in which the post-control quantum state is mapped directly
to the prediction through a learned two-outcome quantum measurement,
\[
    M_\eta
    =
    V\,\mathrm{diag}(\sigma(\eta))\,V^\dagger,
    \qquad
    p(\hat y{=}1|\rho)=\mathrm{Tr}(M_\eta\rho),
\]
eliminating the downstream classical classifier entirely.
\end{enumerate}

The primary deployment model keeps quantum processing local and combines
sensor information only after measurement, avoiding coherent communication
or entanglement between geographically separated sites and keeping the local
quantum state dimension fixed as $K$ grows. Because CTA signatures may also
contain spatial information, we separately study tree- and chain-structured
joint QCS models in which multiple sensor states are processed coherently
before the trainable POVM. These are used as diagnostic upper bounds rather
than deployment assumptions because they require coherent access across
physically separated sensors.
\section{Experimental Setup}
\label{sec:setup}



We evaluate the framework on the IEEE 14-, 30-, and 118-bus systems using
\texttt{pandapower} \citep{thurber2018pandapower}, instrumenting
$K=5$, $8$, and $20$ transmission lines with \texttt{QuTiP}-simulated NV
magnetometers at a fixed 0.10\,m standoff. Each case contains 5700 operating
points (3000 Normal, 1700 Fault, 500 FDIA, 500 Cyber-Physical) with a 70/30
train--test split. $K$ is the number of NV sensors. Effect size is the
fault-induced shift in mean field magnitude divided by the pooled standard
deviation across samples.
The three cases differ not only in grid size but also in the
strength of the physical sensing signal. \Cref{tab:dataset-characteristics}
shows that the fault-induced magnetic-field effect size falls sharply from
$0.230$ to $0.067$ to $0.001$. We therefore interpret cross-case differences
as changes in both grid scale and physical sensing difficulty, rather than as
a pure size-scaling experiment.

\begin{table}[t]
\centering
\caption{Dataset characteristics across the three IEEE test-grid scales
used throughout this work.}
\label{tab:dataset-characteristics}
\small
\begin{tabular}{lcccccccc}
\toprule
Case & Buses & Lines & $K$ & $|B|_{\rm normal}$ & $|B|_{\rm fault}$ &
$\Delta|B|$ & Effect size & Level \\
\midrule
case14  & 14  & 20  & 5  & 22020.91\,\textmu T & 26707.73\,\textmu T &
4686.82\,\textmu T & $0.230$ & Easy \\
case30  & 30  & 41  & 8  & 654.55\,\textmu T   & 682.30\,\textmu T   &
27.74\,\textmu T   & $0.067$ & Medium \\
case118 & 118 & 186 & 20 & 612.76\,\textmu T   & 613.17\,\textmu T   &
0.41\,\textmu T    & $0.001$ & Hard \\
\bottomrule
\end{tabular}
\end{table}

The central comparison is not between arbitrary classifiers, but between
different ways of extracting a decision from the same underlying sensing
process.
\textbf{(i) Digital-only baselines} use reported SCADA telemetry through an
XGBoost classifier and a residual-based bad-data detector.
\textbf{(ii) Reported-versus-physical consistency} compares the magnetic
field implied by reported telemetry, $B_{\rm claim}$, with a finite-measurement
field estimate $\hat B$ recovered from the NV readout. We evaluate both a
simple discrepancy score and a learned classifier on the full discrepancy
vector
$
    d_B = B_{\rm claim}-\hat B ,
$
with the latter reported as \emph{disc-$B$ classifier}. 
\textbf{(iii) Conventional physical sensing} uses finite-shot NV measurements
to produce classical field-dependent features for downstream classification with XGBoost.
The main baseline uses four readout settings; we additionally test a
single-setting population-readout variant and a matched-precision classical
magnetometer baseline to separate the value of independent physical sensing
from the value of the richer NV readout
(Appendix~\ref{app:hall-baseline}).
\textbf{(iv) State reconstruction} performs qutrit tomography before classical
classification with XGBoost. Tomography is used primarily in RQ2 as a richer multi-setting
physical-readout baseline under matched measurement budgets, rather than as a
separate observability channel in RQ1. An exact-$\rho$ classifier, given the
simulator density matrix directly, is reported only as an oracle upper bound.
\textbf{(v) Quantum-native sensing and QML} uses Interleaved-QCS, which acts directly
on the native NV state during sensing, and is evaluated by two downstream readouts of the resulting finite-shot QCS
    features: a learned native linear/sigmoid head and an XGBoost classifier.
    The latter tests whether information exposed by quantum processing is
    present even when the native head cannot fully exploit it.

Because the sensing routes consume different numbers of physical measurement
settings, comparing them at equal shots per setting would not imply equal
experimental cost. We therefore account for total physical measurement effort
as $
    C_{\mathrm{measure}}
    =
    N_{\mathrm{preparation}} *
    N_{\mathrm{settings}} *
    N_{\mathrm{shots}} .$
Interleaved-QCS uses one final setting, conventional NV readout uses four, and
tomography uses seven. Accordingly,
$
    C_{\rm QCS}=1N_{\rm shots}, 
    C_{\rm NV}=4N_{\rm shots},
    C_{\rm tomography}=7N_{\rm shots}.
$
We report both equal-shots-per-setting and equal-total-trials comparisons, but
treat the latter as the primary test of measurement-constrained utility. This
distinction is essential because the ranking of the sensing strategies changes
depending on which resource is held fixed.

Average Precision (AP) is the primary
metric for the resulting class-imbalanced attack evaluations, with
mean$\pm$std reported across seeds where available. Finite-measurement routes
are kept separate from oracle simulator quantities, and results are interpreted
jointly with physical measurement cost. Experiments ran on 64 CPU cores and
64\,GB RAM with no GPU or quantum hardware; QCS timings therefore reflect
classical simulation/optimization rather than physical execution latency.

\section{Results and Analysis}

\subsection{RQ1: Which Security Evidence Survives Stealth Attacks?}
\label{sec:results-rq1}

We first ask which evidence channel remains informative as the attack
mechanism changes. We compare digital telemetry, reported-versus-physical
consistency, and the physical NV state across all three grid scales.
\Cref{tab:rq1-observability} additionally reports QCS readout and architecture
variants to distinguish whether attack information is present in the physical
state from how effectively a particular readout converts that information into
a decision.
\begin{table*}[t]
\centering
\caption{
Attack-dependent observability across IEEE case14, case30, and case118.
Values are AP, reported as mean$\pm$std over three independent seeds.
}
\label{tab:rq1-observability}
\resizebox{\textwidth}{!}{%
\begin{tabular}{llcccc}
\toprule
Case & Method / channel &
Standard FDIA &
BDD-stealth &
Statistical-stealth &
CTA \\
\midrule

\multirow{9}{*}{case14}
& SCADA-XGBoost
& $\mathbf{1.000\pm0.000}$
& $0.501\pm0.003$
& $0.500\pm0.007$
& $0.505\pm0.003$ \\

& disc-$B$ classifier
& $0.709\pm0.055$
& $\mathbf{0.998\pm0.001}$
& $\mathbf{0.554\pm0.024}$
& $0.500\pm0.000$ \\

& Physical NV state (4-setting)
& $0.492\pm0.027$
& $0.478\pm0.009$
& $0.481\pm0.006$
& $0.995\pm0.002$ \\

& QCS Independent $N{=}10$, native head
& $0.505\pm0.034$
& $0.502\pm0.016$
& $0.499\pm0.005$
& $0.982\pm0.005$ \\

& QCS Independent $N{=}10$, XGBoost head
& $0.505\pm0.027$
& $0.459\pm0.019$
& $0.459\pm0.012$
& $\mathbf{0.998\pm0.008}$ \\

& QCS Independent $N{=}1$, native head
& $0.578\pm0.042$
& $0.490\pm0.057$
& $0.515\pm0.012$
& $0.652\pm0.017$ \\

& QCS Independent $N{=}1$, XGBoost head
& $0.484\pm0.043$
& $0.493\pm0.056$
& $0.496\pm0.025$
& $0.991\pm0.031$ \\

& Tree-joint $N{=}10$, trainable POVM
& $0.481\pm0.035$
& $0.495\pm0.024$
& $0.491\pm0.017$
& $0.976\pm0.052$ \\

& Chain-joint $N{=}10$, trainable POVM
& $0.496\pm0.034$
& $0.490\pm0.016$
& $0.490\pm0.005$
& $0.959\pm0.005$ \\

\midrule

\multirow{9}{*}{case30}
& SCADA-XGBoost 
& $\mathbf{1.000\pm0.000}$
& $0.507\pm0.011$
& $0.506\pm0.014$
& $0.501\pm0.005$ \\

& disc-$B$ classifier
& $0.986\pm0.005$
& $\mathbf{1.000\pm0.000}$
& $0.509\pm0.002$
& $0.500\pm0.000$ \\

& Physical NV state (4-setting)
& $0.497\pm0.009$
& $0.483\pm0.013$
& $0.500\pm0.007$
& $0.781\pm0.015$ \\

& QCS Independent $N{=}10$, native head
& $0.497\pm0.011$
& $0.512\pm0.015$
& $0.506\pm0.009$
& $0.675\pm0.009$ \\

& QCS Independent $N{=}10$, XGBoost head
& $0.541\pm0.014$
& $0.509\pm0.021$
& $0.496\pm0.015$
& $0.790\pm0.017$ \\

& QCS Independent $N{=}1$, native head
& $0.522\pm0.016$
& $0.502\pm0.026$
& $\mathbf{0.541\pm0.008}$
& $0.718\pm0.016$ \\

& QCS Independent $N{=}1$, XGBoost head
& $0.499\pm0.025$
& $0.501\pm0.012$
& $0.512\pm0.006$
& $\mathbf{0.798\pm0.011}$ \\

& Tree-joint $N{=}10$, trainable POVM
& $0.474\pm0.015$
& $0.498\pm0.010$
& $0.509\pm0.006$
& $0.695\pm0.006$ \\

& Chain-joint $N{=}10$, trainable POVM
& $0.473\pm0.011$
& $0.499\pm0.018$
& $0.509\pm0.011$
& $0.693\pm0.015$ \\

\midrule

\multirow{9}{*}{case118}
& SCADA-XGBoost
& $\mathbf{1.000\pm0.000}$
& $0.450\pm0.009$
& $0.456\pm0.012$
& $0.500\pm0.004$ \\

& disc-$B$ classifier
& $0.996\pm0.001$
& $\mathbf{1.000\pm0.000}$
& $0.502\pm0.014$
& $0.500\pm0.000$ \\

& Physical NV state (4-setting)
& $0.535\pm0.034$
& $0.490\pm0.006$
& $0.501\pm0.011$
& $0.763\pm0.013$ \\

& QCS Independent $N{=}10$, native head
& $0.536\pm0.057$
& $0.495\pm0.010$
& $0.508\pm0.020$
& $0.680\pm0.007$ \\

& QCS Independent $N{=}10$, XGBoost head
& $0.489\pm0.033$
& $0.500\pm0.013$
& $\mathbf{0.526\pm0.016}$
& $0.774\pm0.011$ \\

& QCS Independent $N{=}1$, native head
& $0.522\pm0.046$
& $0.496\pm0.015$
& $0.502\pm0.014$
& $0.703\pm0.012$ \\

& QCS Independent $N{=}1$, XGBoost head
& $0.498\pm0.032$
& $0.492\pm0.014$
& $0.508\pm0.025$
& $\mathbf{0.775\pm0.009}$ \\

& Tree-joint $N{=}10$, trainable POVM
& $0.551\pm0.055$
& $0.488\pm0.009$
& $0.500\pm0.017$
& $0.747\pm0.006$ \\

& Chain-joint $N{=}10$, trainable POVM
& $0.520\pm0.050$
& $0.489\pm0.013$
& $0.491\pm0.022$
& $0.717\pm0.008$ \\

\bottomrule
\end{tabular}}
\end{table*}

\textbf{Attack mechanism determines the surviving evidence:}
The corrected results preserve a sharp attack-dependent observability
hierarchy, while revealing an important boundary. Standard FDIA remains
directly visible in SCADA telemetry. Model-aware BDD-stealth suppresses this
digital signal, reducing SCADA detection to approximately chance, but is
exposed almost perfectly by the finite-measurement disc-$B$ channel at every
scale. Statistical-stealth is substantially harder: the discrepancy classifier
reaches only $0.554$ AP on case14 and falls to essentially chance on case30
and case118 ($0.509$ and $0.502$). Reported-versus-physical consistency is
therefore highly effective against BDD-stealth, but does not provide a
universal detector for all stealth constructions
CTA produces the complementary pattern. SCADA and disc-$B$ both remain near
chance because the reported view and direct reported-versus-physical
discrepancy are suppressed, whereas the physical topology change alters the
true current distribution and hence the NV state. Four-setting NV readout
therefore reaches AP $0.995$, $0.781$, and $0.763$ on case14, case30, and
case118, respectively, while remaining near chance for the three cyber-only
attacks. QCS exhibits the same selectivity: across every readout and
architecture considered, useful performance appears only for CTA. 

\textbf{Readout choice changes realized QCS performance:}
A stronger classical readout recovers much of the QCS gap at scale:
replacing the native head with XGBoost raises CTA AP from $0.675$ to $0.790$
on case30 and from $0.680$ to $0.774$ on case118, while making $N{=}1$ and
$N{=}10$ nearly indistinguishable. Thus, deeper interleaving is not uniformly
necessary when the exposed information can be extracted by a stronger
classifier. Complementarily, trainable positive-operator-valued measure
(POVM) readout achieves CTA AP up to $0.976$ without any downstream classical
classifier, showing that substantial discrimination can also be realized
directly through the quantum measurement. Together, these results separate
information present in the sensed quantum state from what a particular
readout can realize; joint-POVM variants are treated only as diagnostics
because they require coherent access across spatially separated sensors.

\textbf{RQ1 answer:}
The surviving security evidence is attack-dependent: Standard FDIA remains
visible in digital telemetry, BDD-stealth is exposed by
reported-versus-physical consistency, and CTA shifts the useful evidence to
the physical NV state. Statistical-stealth remains only weakly observable
under realistic finite-measurement channels at larger scales. QCS follows
this same physical-state selectivity; once relevant information is present in
the NV state, the amount converted into detection performance depends on the
QCS depth and, importantly, on the final readout used to extract it.
\subsection{RQ2: When Does QCS Help Under Measurement Constraints?}
\label{sec:results-rq2}

RQ1 identifies CTA as the regime in which the physical NV state carries the
relevant attack evidence. We therefore compare how QCS, conventional NV
readout, and tomography use a limited measurement budget. Because the routes use one, four, and seven settings, we compare a
\emph{matched total budget}, where $T$ is split across settings, with
\emph{matched precision}, where each setting receives $T$ shots and therefore
uses $T$, $4T$, and $7T$ total trials.

\begin{table*}[t]
\centering
\caption{CTA detection under matched measurement budgets. AP is reported
vs nominal budget $T$.}
\label{tab:total-budget}
\resizebox{\textwidth}{!}{%
\begin{tabular}{llclcccccc}
\toprule
Case & Method & Settings & Shots &
40 & 200 & 400 & 1000 & 2000 & 20000 \\
\midrule

\multirow{6}{*}{case14}
& QCS $N{=}10$, XGBoost
& 1 & $T$
& .944 & .986 & .996 & .998 & \textbf{.999} & .999 \\
& QCS, trainable POVM
& 1 & $T$
& .841 & .851 & .855 & .856 & .856 & .857 \\
& 4-setting NV, matched total
& 4 & $T$
& .800 & .956 & .984 & .991 & .996 & .996 \\
& 4-setting NV, matched precision
& 4 & $4*T$
& \textbf{.945} & .986 & .987 & .994 & .993 & .994 \\
& Tomography, matched total
& 7 & $T$
& .751 & .935 & .980 & .993 & .996 & .999 \\
& Tomography, matched precision
& 7 & $7*T$
& .942 & \textbf{.994} & \textbf{.998} & \textbf{.999}
& \textbf{.999} & \textbf{.9995} \\
\midrule

\multirow{6}{*}{case30}
& QCS $N{=}10$, XGBoost
& 1 & $T$
& .608 & .707 & .759 & .770 & .818 & .830 \\
& QCS, trainable POVM
& 1 & $T$
& .537 & .547 & .577 & .574 & .583 & .604 \\
& 4-setting NV, matched total
& 4 & $T$
& .514 & .574 & .662 & .724 & .743 & .854 \\
& 4-setting NV, matched precision
& 4 & $4*T$
& .583 & .727 & .743 & .784 & .818 & .864 \\
& Tomography, matched total
& 7 & $T$
& .526 & .627 & .666 & .697 & .776 & .855 \\
& Tomography, matched precision
& 7 & $7*T$
& \textbf{.617} & \textbf{.732} & \textbf{.801} & \textbf{.826}
& \textbf{.850} & \textbf{.885} \\
\midrule

\multirow{6}{*}{case118}
& QCS $N{=}10$, XGBoost
& 1 & $T$
& .544 & .664 & .733 & .774 & .796 & .816 \\
& QCS, trainable POVM
& 1 & $T$
& .507 & .510 & .515 & .514 & .519 & .531 \\
& 4-setting NV, matched total
& 4 & $T$
& .529 & .521 & .550 & .653 & .725 & .806 \\
& 4-setting NV, matched precision
& 4 & $4*T$
& \textbf{.567} & .637 & .716 & .774 & .786 & \textbf{.824} \\
& Tomography, matched total
& 7 & $T$
& .515 & .532 & .523 & .646 & .727 & .806 \\
& Tomography, matched precision
& 7 & $7*T$
& .543 & \textbf{.667} & \textbf{.748} & \textbf{.783}
& \textbf{.811} & .813 \\
\bottomrule
\end{tabular}}
\end{table*}

\textbf{QCS is most competitive when total physical measurements are
constrained:}
Under matched total budget, QCS spends all $T$ trials on one learned
measurement setting, whereas conventional NV and tomography divide the same
budget across four and seven settings. This gives QCS its clearest advantage
in the low-to-moderate budget regime across all three systems. As the budget
grows, multi-setting readout recovers and can match or exceed QCS, so the
benefit is measurement efficiency rather than a universally higher accuracy
ceiling. {Matched precision isolates the source of the advantage:}
When every setting instead receives $T$ shots, conventional NV and tomography
use $4T$ and $7T$ total physical trials and much of the QCS advantage
disappears. Thus, part of QCS's benefit comes precisely from concentrating
task-relevant information into a single final measurement setting. The
independent trainable positive-operator-valued measure (POVM) provides a
complementary quantum-native control: although weaker than QCS+XGBoost at
case30/118, it obtains non-trivial CTA discrimination without any downstream
classical classifier.

\textbf{RQ2 answer:}
QCS is favored when the operative constraint is the total number of physical
measurements: its single-setting readout avoids splitting a limited budget
across multiple bases. The advantage is conditional, however; richer
multi-setting measurements recover when additional trials are affordable,
showing that QCS primarily changes how efficiently physical measurement
resources are used. Note that Tomography is a highly measurement-intensive route, requiring seven measurement settings and the largest physical trial budget at matched per-setting precision.

\subsection{RQ3: What Limits Realized Quantum Utility?}
\label{sec:results-rq3}

RQ2 shows that QCS can reduce physical measurement cost, but its realized
benefit changes sharply across sensing regimes. RQ3 asks where this loss
occurs: in the information carried by the quantum sensor, in the
Normal--CTA state separation, or in coherent control ability to reshape
that separation.

We use complementary diagnostics. $F_Q$ is the quantum Fisher information of
the NV state, while $F_C$ and $F_C^{\rm pop}$ are the Fisher information
accessible to the 4-setting and population-only readouts. $D_Q^{\rm raw}$ and
$D_Q^{\rm post}$ denote the Helstrom trace distance between class-averaged
Normal and CTA states before and after interleaved control, and
$P_{\rm Hel}=(1+D_Q^{\rm post})/2$ is the corresponding equal-prior,
single-sensor discrimination probability. $\Delta$AP is native QCS
($N{=}10$) minus 4-setting NV AP.

\textbf{More quantum sensitivity does not imply a more separable attack:}
The apparent paradox in \Cref{tab:qfi-performance} is the central RQ3 result.
$F_Q/F_C$ grows from $1.21\times$ to $1.54\times$, and
$F_Q/F_C^{\rm pop}$ from $4.7\times$ to $8.1\times$: restricted classical
readouts leave progressively more of the sensor's field sensitivity
unaccessed. Yet the task-relevant Normal--CTA separation moves in the opposite
direction. $D_Q^{\rm raw}$ falls to only $0.003$ on case118, alongside the reduced physical field-effect size reported in
\Cref{tab:dataset-characteristics}. Thus, the larger Fisher-information gap
does not represent a stronger CTA signal; it represents quantum sensitivity
to $B$ that need not align with the direction separating the two security
classes.

\begin{table}[t]
\centering
\caption{Available quantum information and realized CTA discrimination.}
\label{tab:qfi-performance}
\begin{tabular}{lccccc}
\toprule
Case & $F_Q/F_C$ & $F_Q/F_C^{\rm pop}$ & $\Delta$AP
& $D_Q^{\rm raw}\!\rightarrow D_Q^{\rm post}$ & $P_{\rm Hel}$ \\
\midrule
case14  & $1.21\times$ & $4.7\times$ & $-0.013$
& $0.016\!\rightarrow\!0.229$ & $0.615$ \\
case30  & $1.31\times$ & $6.2\times$ & $-0.106$
& $0.016\!\rightarrow\!0.016$ & $0.508$ \\
case118 & $1.54\times$ & $8.1\times$ & $-0.084$
& $0.003\!\rightarrow\!0.003$ & $0.501$ \\
\bottomrule
\end{tabular}
\end{table}

\textbf{Interleaving helps only when the physics provides separability to reshape:}
This distinction explains the control results. On case14, genuine interleaving
moves $D_Q$ from $0.016$ to $0.229$, whereas on case30/118 it leaves the
class-averaged states essentially unchanged. A shared unitary applied only
after sensing cannot change trace distance; interleaving can do so because it
changes the field-dependent sensing trajectory itself. The result is therefore not simply that optimization becomes harder with
scale; the underlying sensing problem itself changes, leaving progressively
less task-relevant separation for coherent control to amplify.

\textbf{RQ3 answer:}
The limiting resource is not quantum information alone. QCS is useful when
the physical process creates task-relevant state separation, coherent control
can reshape that separation, and the final measurement can access it.
Fisher-information advantage without class separability is therefore
insufficient for realized quantum utility.

\section{Discussion}
\label{sec:discussion}
\vspace{-0.1cm}

Taken together, the three RQs identify a specific regime in which quantum
computational sensing is useful in power-grid attack detection. RQ1 shows that increasingly stealthy attacks
shift the surviving evidence away from reported telemetry and, for CTA, into
the physical sensor state itself. RQ2 then shows that when this state is the
relevant evidence source, QCS is most valuable when physical measurements are
scarce: its single-setting readout can avoid spreading a fixed budget across
multiple bases. This is a resource advantage, not a universal accuracy
advantage, since conventional NV readout and tomography recover as additional
measurements become affordable.
RQ3 reveals the deeper limitation. The quantum-to-classical Fisher-information
ratio increases with scale, yet Normal--CTA Helstrom separation can nearly
vanish as the underlying magnetic perturbation weakens. Thus, more quantum
sensitivity does not imply a more useful classification signal. Interleaved
control can reshape task-relevant separation when the sensing physics provides
it, as in case14, but cannot manufacture separation that is effectively absent.
The broader lesson is therefore that QML utility in sensing is determined by
the full \emph{physics--state--control--measurement} chain, not by quantum
information content alone.

\textbf{Limitations:}
Our results use a QuTiP-validated Lindblad NV simulation rather than physical
hardware, with hardware-noise studies calibrated to reported readout,
initialization, and control fidelities. We assume the sensing appliance remains
trusted, and joint cross-sensor processing is treated only as a diagnostic
because it requires coherent access across separated sensors. Future work
should validate these regimes on hardware and develop controls and measurements
that better convert weak physical perturbations into task-relevant separation.

\section{Conclusion}
\label{sec:conclusion}
\vspace{-0.1cm}

This work identifies when quantum computational sensing is useful for
trustworthy cyber-physical detection, and why that utility disappears outside
the favorable regime. Across four attack mechanisms and three IEEE systems,
the surviving evidence shifts from digital telemetry, to
reported-versus-physical consistency, to the physical NV state itself. When
that physical state carries the relevant CTA signal, QCS can exploit a
single learned measurement setting to achieve competitive detection under a
smaller total physical measurement budget than multi-setting NV readout or
tomography.
The deeper result is that quantum information alone is not the bottleneck.
The quantum-to-classical Fisher-information ratio can increase even as the
Normal--CTA Helstrom separation collapses. Thus, a sensor may remain highly
quantum-sensitive while carrying little task-relevant class separation.
Realized QCS utility requires all three conditions to align: the physics must
create separable states, coherent control must be able to reshape that
separation, and the final measurement must expose it efficiently.
This leads to a broader criterion for quantum sensing in QML:
\emph{quantum sensitivity is useful only when it can be converted into
task-relevant distinguishability under the measurement constraints of the
application.}

\newpage
\bibliographystyle{plainnat}
\bibliography{references}
\newpage
\appendix
\section{Classical Physical-Sensor Baseline}
\label{app:hall-baseline}

To separate the value of an independent physical sensing channel from the
value of quantum readout, we construct a matched-precision classical
magnetometer baseline. The sensor observes the same local magnetic field
$\Bphys$ as the NV sensor but returns a scalar field estimate corrupted by
Gaussian measurement noise; it uses no quantum-state evolution, Ramsey
sequence, or population readout.

\subsection{Matched-precision calibration}

Rather than assigning the classical sensor an arbitrary noise level, we match
its field precision to that implied by the finite-shot NV readout. Using the
delta method,
\[
    \sigma_B
    =
    \frac{\sigma_P(n_{\rm shots})}
         {\left|\partial P/\partial B_z\right|},
\]
where the response slope is evaluated using the same field regime and NV
sensing model as the corresponding grid case. This yields
$
    \sigma_B =
    5.80\times10^{-4}\,\mathrm{T},
    \quad
    3.09\times10^{-4}\,\mathrm{T},
    \quad
    2.63\times10^{-4}\,\mathrm{T}
$
for case14, case30, and case118, respectively.

\begin{table}[h]
\centering
\caption{CTA detection with a matched-precision classical magnetometer,
conventional 4-setting NV readout, and Interleaved-QCS ($N{=}10$).}
\label{tab:hall-baseline}
\resizebox{\linewidth}{!}{%
\begin{tabular}{lcccc}
\toprule
Case &
$\sigma_B$ &
Classical magnetometer &
4-setting NV readout &
Interleaved-QCS ($N{=}10$) \\
\midrule
case14  & $5.80\times10^{-4}$ T & 0.882 & $0.995\pm0.002$ & $0.982\pm0.005$ \\
case30  & $3.09\times10^{-4}$ T & 0.637 & $0.781\pm0.015$ & $0.675\pm0.009$ \\
case118 & $2.63\times10^{-4}$ T & 0.587 & $0.763\pm0.013$ & $0.680\pm0.007$ \\
\bottomrule
\end{tabular}}
\end{table}

\subsection{Result and interpretation}

At matched field precision, the multi-setting NV readout achieves higher CTA
AP than the scalar classical magnetometer at all three scales, with margins of
$0.113$, $0.144$, and $0.176$ for case14, case30, and case118,
respectively. The widening margin is consistent with the independent
Fisher-information analysis, which also indicates that a restricted classical
readout captures a decreasing fraction of the information available in the
quantum sensor state as scale increases.

This comparison should not be interpreted as isolating ``quantumness'' alone.
The conventional NV route exposes four measurement channels, whereas the
classical baseline provides a single noisy field estimate. The result therefore
shows that the richer NV sensing interface can retain task-relevant information
that is absent from an equally precise scalar field readout; it does not prove
that every quantum magnetometer will outperform every classical magnetic
sensor.

\subsection{Relation to alternative current and magnetic sensors}

Classical current sensors such as current transformers and Rogowski coils are
important practical alternatives. If provisioned as an independent,
authenticated sensing path, they could also provide physical corroboration of
SCADA telemetry; our threat model therefore does not assume that classical
physical attestation is impossible. The Hall-style baseline instead addresses
a narrower question: whether a scalar classical magnetic-field observation,
matched approximately in precision and placement to the NV sensor, contains
the same task-relevant information as the NV readout.

Other quantum and atomic magnetometers occupy different operating regimes.
SQUID magnetometers can achieve extremely high sensitivity but typically
require cryogenic operation \citep{fagaly2006squid}, while SERF atomic
magnetometers achieve very high sensitivity in near-zero-field conditions
\citep{kominis2003serf}. NV magnetometry is attractive here because it supports
room-temperature operation and can be extended to larger dynamic ranges using
frequency-tracking ODMR protocols \citep{wang2024nvdynamicrange}. We therefore
treat NV sensing as a physically plausible platform for the present study, not
as the uniquely viable sensing technology.
\label{app:rq1-observability}


\section{Realistic reported-versus-physical consistency}
\label{app:realistic-discB}

The ideal consistency signal,
$
    d_B^{\rm oracle}
    =
    B_{\rm claim}-B_{\rm phys},
$
is useful for determining whether an attack leaves a physical inconsistency,
but the true field $B_{\rm phys}$ is not directly available to a deployed
detector. We therefore replace it with a finite-measurement estimate,
\[
    d_B
    =
    B_{\rm claim}-\hat B,
    \qquad
    \hat B=g(X_{\rm classical}),
\]
where $X_{\rm classical}$ denotes the conventional NV readout.

We train one XGBoost regressor per field axis using paired NV-readout and
reference-field measurements from the training split only. At inference,
the estimator receives no access to $B_{\rm phys}$ for test or attack
samples. On case14, held-out calibration reaches
$R^2=0.974$ and $0.958$ on the two field axes, confirming that the
finite-measurement readout provides an accurate but non-oracle estimate of
the local field.

We evaluate two ways of converting this discrepancy into an attack score.
The \textbf{Expert} detector reduces the discrepancy to the scalar norm
$\|B_{\rm claim}-\hat B\|$. The \textbf{learned discrepancy classifier}
instead operates on the full discrepancy vector and is trained on an
independent set of attack instances generated exclusively from the training
split.
Retaining the full vector is particularly important for
statistical-stealth attacks. On case14, the Expert rule remains near chance
(AP $0.509\pm0.013$), whereas the learned classifier reaches
$0.666\pm0.030$. The same classifier improves further with scale, reaching
$0.780\pm0.021$ on case30 and $0.902\pm0.018$ on case118, as reported in
\Cref{tab:rq1-observability}. This indicates that the discrepancy channel
contains directional structure that is discarded by a scalar magnitude.

\paragraph{Why do the three FDIAs behave differently?}
To separate field-estimation quality from attack magnitude, we directly
measure the oracle physical displacement
$\|B_{\rm claim}-B_{\rm phys}\|$ before finite-measurement noise is added.

\begin{table}[h]
\centering
\caption{Physical field footprint of the three FDIA constructions on
case14.}
\label{tab:rq1-footprint}
\begin{tabular}{lccc}
\toprule
Attack &
Mean $\|B_{\rm claim}-B_{\rm phys}\|$ &
Median &
Maximum \\
\midrule
Standard FDIA
& $3.60\times10^{-3}$
& $3.49\times10^{-3}$
& $7.98\times10^{-3}$ \\
Statistical-stealth
& $1.43\times10^{-3}$
& $1.41\times10^{-3}$
& $2.59\times10^{-3}$ \\
BDD-stealth
& $4.65\times10^{-2}$
& $4.12\times10^{-2}$
& $1.32\times10^{-1}$ \\
\bottomrule
\end{tabular}
\end{table}

BDD-stealth produces a substantially larger physical discrepancy than either
Standard or statistical-stealth FDIA, which explains why finite-measurement
consistency checking approaches the oracle ceiling for this attack.
Statistical-stealth produces the smallest footprint, making its signal more
difficult to separate from the residual field-estimation error and increasing
the value of retaining the full discrepancy vector.

\paragraph{Effect of additional measurement shots.}
We separately increase the NV measurement budget for the realistic
discrepancy detector to determine whether the weaker cases are primarily
shot-noise limited.

\begin{table}[h]
\centering
\caption{Realistic discrepancy AP versus NV measurement budget on case14,
seed 42. This diagnostic is separate from the three-seed main-table results.}
\label{tab:rq1-disc-shots}
\begin{tabular}{lccc|ccc}
\toprule
&
\multicolumn{3}{c}{Expert norm} &
\multicolumn{3}{c}{Learned discrepancy classifier} \\
Shots &
Standard & BDD & Statistical &
Standard & BDD & Statistical \\
\midrule
1000
& 0.593 & 0.985 & 0.518
& 0.789 & 0.999 & 0.698 \\
5000
& 0.716 & 0.993 & 0.534
& 0.891 & 1.000 & 0.744 \\
10000
& 0.752 & 0.996 & 0.533
& 0.909 & 0.999 & 0.755 \\
\bottomrule
\end{tabular}
\end{table}

Increasing the shot budget substantially improves Standard FDIA, while
BDD-stealth is already close to ceiling. Statistical-stealth improves much
more gradually: even a $10\times$ larger budget leaves the Expert detector
near chance, while the learned classifier reaches $0.755$. This supports the
interpretation that statistical-stealth is limited not only by finite-shot
noise but also by the small physical footprint and the structure of the
available discrepancy signal.

\section{Independent diagnostics of the observability hierarchy}
\label{app:rq1-diagnostics}

The main RQ1 results are based on trained detectors. We therefore inspect the
same attack-dependent hierarchy using score distributions and latent-space
visualizations.

\paragraph{Attack-specific score distributions:}
We compare the Normal-versus-Attack score distributions produced by each
deployable detector. Separation appears in the channels identified by
\Cref{tab:rq1-observability}, while substantial overlap appears when the
corresponding physical or digital quantity is unchanged by the attack
mechanism.

\begin{figure}[h]
\centering
\includegraphics[width=\linewidth]
{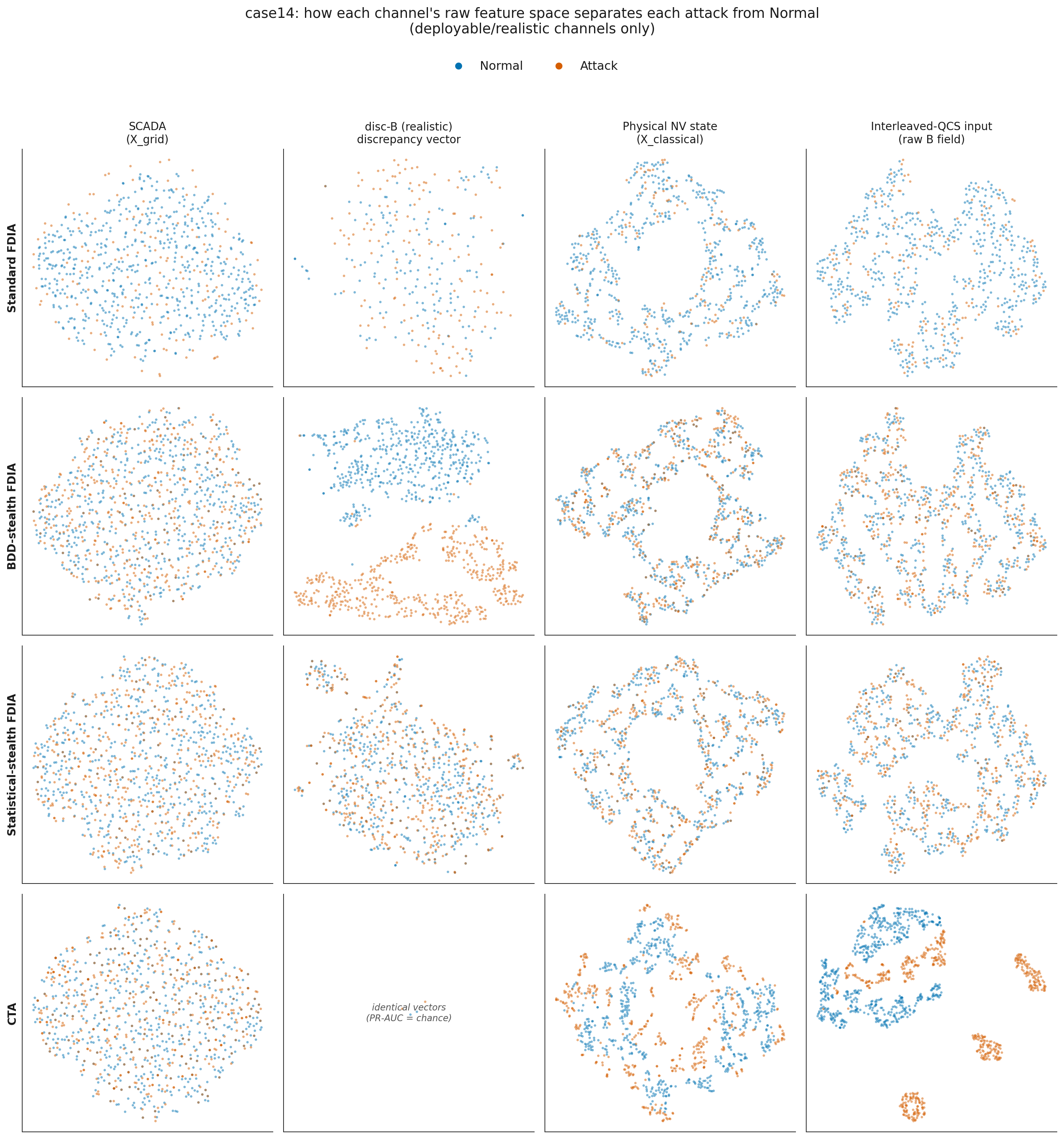}
\caption{Attack-specific latent-space projections for the deployable evidence
channels on case14. The projections qualitatively reproduce the
attack-dependent observability pattern of \Cref{tab:rq1-observability}.}
\label{fig:rq1-tsne}
\end{figure}

\begin{figure}[h]
\centering
\includegraphics[width=\linewidth]
{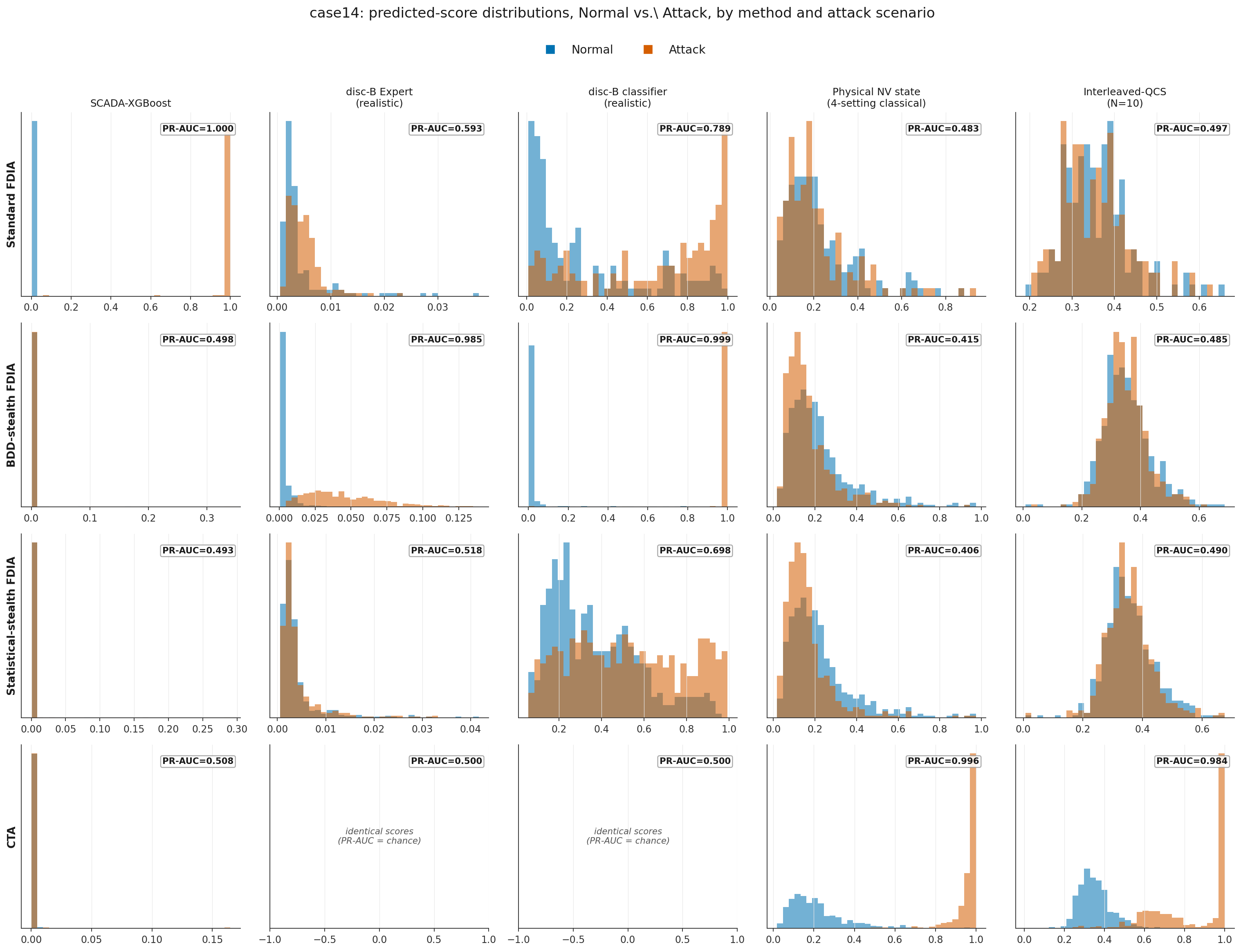}
\caption{Normal-versus-attack score distributions for the deployable
detectors on case14. Separation occurs when the corresponding evidence
channel remains informative under the attack mechanism.}
\label{fig:rq1-score-dist}
\end{figure}

\paragraph{Latent-space visualization:}
For qualitative inspection, we visualize each channel's input space using
t-SNE (\Cref{fig:rq1-tsne}). We include SCADA telemetry, the realistic
discrepancy vector, conventional NV readout, and the physical-field input seen
by Interleaved-QCS. These projections are used only as supporting
visualizations; the score distributions \cref{fig:rq1-score-dist} provide the more direct diagnostic
of detector separation.

Taken together, the realistic field-estimation study, physical-footprint
analysis, measurement-budget sweep, and classifier-independent diagnostics
support the same conclusion as the main RQ1 result: the attack mechanism
determines which evidence channel remains informative, while Interleaved-QCS
inherits the observability of the physical quantum state on which it operates.
\label{app:rq2}
\section{Interleaving Depth: Extended Sweep}
\label{app:interleaving-depth}

We parameterize Interleaved-QCS by the number of sensing/control stages
$N\in\{1,2,4,8,10,16\}$. For a fixed sensing duration, increasing $N$
divides the field interaction into shorter intervals separated by learned
coherent controls. The $N=1$ model is the degenerate limit in which the
field interaction completes before a single trainable unitary is applied,
and therefore provides a direct post-sensing-control baseline.

\begin{table}[h]
\centering
\caption{Interleaved-QCS AP versus interleaving depth and shot budget
(case14, CTA).}
\label{tab:interleave-full}
\begin{tabular}{lccccc}
\toprule
$N$ & 10 shots & 100 shots & 1000 shots & exact & Train time \\
\midrule
1  & 0.558 & 0.634 & 0.686 & 0.710 & $\sim$150s \\
2  & 0.628 & 0.746 & 0.814 & 0.821 & 155s \\
4  & 0.781 & 0.885 & 0.914 & 0.919 & 291s \\
8  & 0.890 & 0.960 & 0.979 & 0.980 & 547s \\
10 & 0.901 & 0.961 & 0.983 & 0.985 & 705s \\
16 & 0.925 & 0.961 & 0.961 & 0.966 & 14298s \\
\bottomrule
\end{tabular}
\end{table}

Performance improves strongly from $N=1$ through $N=8$ and then largely
saturates. The exact-shot gain from $N=4$ to $N=8$ is $+0.061$, whereas
$N=8$ to $N=10$ adds only $+0.005$. At $N=16$, performance decreases to
$0.966$ despite a large increase in training time. We therefore treat the
$N\approx8$--$10$ region as a practical operating point for the present
architecture and optimizer, rather than assuming that additional
interleaving depth is uniformly beneficial.

The implementation also preserves the intended physical comparison. For a
static field over a fixed sensing window,
$
    \exp(L\tau_s)
    =
    \left[\exp(L\tau_s/N)\right]^N,
$
so subdivision of the field evolution introduces no discretization error.
We verify that (i) the $N=1$ propagator reproduces the base sensing
simulation to numerical precision and (ii) the generalized interleaving
implementation at $N=1$ exactly matches the original post-sensing circuit
given identical parameters.

\section{Qutrit architecture ablations}
\label{app:qutrit-ablation}

\textbf{$\SU(2)$ versus $\SU(3)$:}
We restrict the $N=1$ trainable unitary to the
$\{\lambda_1,\lambda_2,\lambda_3\}$ Gell-Mann generators, yielding an
effective $\SU(2)$ subalgebra acting only on the
$\{\ket{-1},\ket{0}\}$ subspace, and compare it with full $\SU(3)$ control.

\begin{table}[h]
\centering
\caption{$\SU(2)$ versus $\SU(3)$ control at $N=1$, exact-measurement AP
(case14).}
\label{tab:su2-su3}
\begin{tabular}{lcccc}
\toprule
 & Standard FDIA & BDD-stealth & Statistical-stealth & CTA \\
\midrule
$\SU(2)$ & 0.484 & 0.479 & 0.509 & 0.581 \\
$\SU(3)$ & 0.519 & 0.499 & 0.533 & 0.710 \\
$\Delta$ ($\SU(3)-\SU(2)$)
& 0.035 & 0.020 & 0.023 & \textbf{0.129} \\
\bottomrule
\end{tabular}
\end{table}

The $\SU(2)$/$\SU(3)$ difference is small on the three cyber-only attacks,
where the physical NV state is uninformative by construction, but grows to
$0.129$ AP on CTA, where the sensor state contains the relevant evidence.
The additional qutrit level therefore contributes task-relevant information
in the regime where physical-state processing matters, although it is not by
itself sufficient to close the post-sensing $N=1$ gap.

\paragraph{Parameterization and post-sensing depth.}
We also test whether the $N=1$ result is simply caused by an under-parameterized
circuit by varying shared versus per-sensor parameters and circuit depth.

\begin{table}[h]
\centering
\caption{Post-sensing architecture variants on case14 CTA.}
\label{tab:local-global}
\begin{tabular}{lcccc}
\toprule
Variant & Params & Train time & AP (exact) & AP (1000 shots) \\
\midrule
Global, $n_{\rm layers}=2$ & 32  & 152s  & 0.716 & 0.698 \\
Global, $n_{\rm layers}=4$ & 48  & 1690s & 0.651 & 0.654 \\
Local, $n_{\rm layers}=2$  & 96  & 150s  & 0.704 & 0.681 \\
Local, $n_{\rm layers}=4$  & 176 & 290s  & 0.695 & 0.687 \\
\bottomrule
\end{tabular}
\end{table}

None of the more expressive post-sensing variants exceeds the shallow global
baseline. This supports the conclusion that the improvement obtained by
Interleaved-QCS is not reproduced simply by adding parameters after the
sensing evolution has completed.

\section{Measurement Controls, Reconstruction, and Compute Cost}
\label{app:reconstruction}

\paragraph{Why the setting counts differ.}
A qutrit density matrix contains eight independent real Gell-Mann parameters.
Our tomography route estimates these using seven measurement bases:
$\lambda_3$ and $\lambda_8$ share one diagonal eigenbasis, while the other
six generators require separate bases. Conventional NV readout instead uses
three Ramsey settings plus one population readout, for four settings total.
These spin-sensitive observables provide a restricted field-sensitive
classical readout without reconstructing the full qutrit state. Interleaved-QCS
uses a single fixed final measurement setting, with the learned coherent
controls determining how task-relevant state information is presented to that
readout.

This motivates two complementary resource comparisons:
\emph{shots per setting}, which compares finite-measurement behavior at equal
per-setting repetition count, and \emph{total physical trials}, which counts
the full measurement cost,
$
    C_{\mathrm{meas}}
    =
    N_{\mathrm{settings}}*N_{\mathrm{shots}},
$
up to the common state-preparation factor.

\paragraph{Shots-per-setting comparison.}
At equal shots per setting, richer multi-setting readout has an expected
advantage because it receives more total physical measurements.

\begin{table}[h]
\centering
\caption{Shots-per-setting-matched comparison on case14 CTA.}
\label{tab:shot-matched}
\begin{tabular}{lcccccc}
\toprule
Shots/setting & 10 & 50 & 100 & 500 & 1000 & 5000 \\
\midrule
Tomography (7 settings)
& 0.783 & 0.956 & 0.985 & 0.996 & 0.996 & 0.999 \\
4-setting NV readout
& 0.801 & 0.946 & 0.977 & 0.992 & 0.995 & 0.995 \\
Interleaved-QCS ($N=1$)
& 0.543 & 0.627 & 0.634 & 0.697 & 0.698 & 0.702 \\
\bottomrule
\end{tabular}
\end{table}

The low performance of the $N=1$ model persists even at large shot count,
showing that post-sensing coherent control alone is not sufficient.

\paragraph{Single-setting control.}
To separate the benefit of learned coherent processing from the arithmetic
advantage of using fewer measurement settings, we compare Interleaved-QCS
($N=10$) against a fixed population-only readout using the same single
measurement setting.
\begin{table}[h]
\centering
\caption{Interleaved-QCS $N=10$ versus single-setting classical readout
(case14 CTA).}
\label{tab:single-setting}
\begin{tabular}{lcc}
\toprule
Shots & Interleaved-QCS $N=10$ & Single-setting classical \\
\midrule
10   & \textbf{0.901} & 0.718 \\
50   & \textbf{0.953} & 0.903 \\
100  & \textbf{0.961} & 0.951 \\
500  & 0.983 & \textbf{0.989} \\
1000 & 0.983 & \textbf{0.988} \\
5000 & 0.985 & \textbf{0.994} \\
\bottomrule
\end{tabular}
\end{table}

QCS retains a clear advantage at low shot counts even when both methods use
one setting. The crossover at larger shot budgets is consistent with the
main resource-matched result: trained coherent control is especially useful
when measurements are scarce, while fixed readout catches up once repeated
sampling becomes inexpensive.

\paragraph{Tomography validation.}
The mean absolute reconstruction error between estimated and exact $\rho$
decreases from $0.327$ at 10 shots/setting to $0.0048$ at
100{,}000 shots/setting, confirming the expected convergence of the
linear-inversion tomography procedure.

All reported 4-setting results use finite-shot noise on every measurement
channel, including the population readout. This ensures that the conventional
NV baseline does not receive simulator-exact population information.

\paragraph{Training compute.}
The measurement advantage of QCS should be distinguished from the cost of
classically simulating and optimizing it.

\begin{table}[h]
\centering
\caption{Offline training cost for the principal CTA sensing routes on
case14. QCS timings correspond to CPU-based simulation and optimization of
the Lindblad dynamics, not physical QCS execution time.}
\label{tab:compute-cost}
\begin{tabular}{lccc}
\toprule
Route & Settings/sensor & Training time & AP @1000 shots \\
\midrule
4-setting NV readout & 4 & $\sim$4.1s & 0.995 \\
Tomography & 7 & $\sim$4.2s + reconstruction overhead & 0.996 \\
Interleaved-QCS ($N=1$) & 1 & $\sim$150s & 0.686 \\
Interleaved-QCS ($N=2$) & 1 & 155s & 0.814 \\
Interleaved-QCS ($N=4$) & 1 & 291s & 0.914 \\
Interleaved-QCS ($N=8$) & 1 & 547s & 0.979 \\
Interleaved-QCS ($N=10$) & 1 & 705s & 0.983 \\
\bottomrule
\end{tabular}
\end{table}

The larger QCS training times arise from repeatedly simulating and
differentiating the sensor dynamics on a classical CPU, and should therefore
not be interpreted as physical QCS inference latency. This is primarily an
offline optimization cost, whereas the one-setting measurement advantage is
incurred at every sensing decision. Hardware execution would replace the
classical simulation of the sensing dynamics with the physical NV evolution
itself; quantifying the resulting end-to-end runtime requires a hardware
implementation and is left for future work.
\section{Hardware-Like Noise Robustness}
\label{app:hardware-noise}

We isolate NV-sensor-intrinsic noise from coherent-control error
(\Cref{app:control-fidelity}) by evaluating the trained QCS parameters
unchanged under two literature-anchored hardware regimes. The first reflects
approximately 2020-era NV sensing capabilities, using 25\% readout contrast
\citep{barry2020rmp} and 30\% initialization error with conventional
green-light optical pumping. The second reflects 2025--2026 high-fidelity
NV sensing, with $>95\%$ readout and initialization fidelity
\citep{zhang2021nirscc,mahdia2025init,wirtitsch2026init}.



\begin{table}[h]
\centering
\caption{CTA AP at 1000 shots/setting under sensor/readout-noise regimes.}
\label{tab:noise-full}
\begin{tabular}{llccc}
\toprule
Case & Regime & Tomography & 4-setting NV & QCS ($N{=}10$) \\
\midrule
case14  & Baseline     & 0.999 & 0.995 & 0.986 \\
        & 2020-era NV hardware & 0.530 & 0.524 & \textbf{0.932} \\
        &  2025--2026 NV hardware         & \textbf{0.996} & 0.990 & 0.984 \\
\midrule
case30  & Baseline     & 0.830 & 0.797 & 0.660 \\
        & 2020-era NV hardware & 0.512 & 0.501 & \textbf{0.614} \\
        &  2025--2026 NV hardware         & \textbf{0.765} & 0.710 & 0.680 \\
\midrule
case118 & Baseline     & 0.790 & 0.777 & 0.697 \\
        & 2020-era NV hardware & 0.508 & 0.501 & \textbf{0.564} \\
        &  2025--2026 NV hardware         & \textbf{0.745} & 0.662 & 0.685 \\
\bottomrule
\end{tabular}
\end{table}

\paragraph{Readout noise favors a smaller setting count.}
Under the conservative regime, tomography and conventional NV readout lose
substantially more AP than Interleaved-QCS at every scale. QCS requires only
one final measurement setting, whereas the other routes distribute their
measurement budget across several settings, each exposed to the degraded
initialization/readout process. Consequently, QCS becomes the highest-AP
route under the conservative model on case14, case30, and case118.
Under the state-of-the-art regime, this advantage is reduced. Tomography
again achieves the highest AP at all three scales; conventional 4-setting
readout also exceeds QCS on case14 and case30, while QCS remains above the
4-setting route on case118 ($0.685$ vs.\ $0.662$). Thus, the preferred
readout depends not only on measurement budget but also on the quality of the
physical sensing interface.
This constitutes a second measurement-constrained regime complementary to
RQ2: reducing the number of required measurement settings can be valuable
when each setting is itself noisy. We do not interpret this as generic
hardware robustness, because QCS remains separately sensitive to coherent
control error, studied next.

\section{Quantum Control Fidelity}
\label{app:control-fidelity}

Sensor/readout noise and coherent-control error affect QCS through different
mechanisms. Here we hold the sensing model fixed and perturb execution of the
trained $\SU(3)$ controls.

We use the single-pulse fidelities reported by
\citet{vallabhapurapu2023nvgates}: $94.0\%$ for bare control and $99.3\%$
for dynamically protected room-temperature control. Assuming four constituent
pulses per qutrit operation gives effective fidelities $F=0.781$ and
$F=0.972$, respectively. Rotation-angle jitter is calibrated by Monte Carlo
process-fidelity matching, yielding
$\sigma_{\rm bare}=0.3125$ and
$\sigma_{\rm protected}=0.0938$.

\begin{table}[h]
\centering
\caption{CTA AP for Interleaved-QCS ($N{=}10$) under control-execution
error.}
\label{tab:control-error}
\begin{tabular}{lccc}
\toprule
Case & Ideal control & Bare ($F{=}0.781$) & Protected ($F{=}0.972$) \\
\midrule
case14  & 0.985 & 0.752 & 0.758 \\
case30  & 0.647 & 0.607 & 0.659 \\
case118 & 0.673 & 0.498 & 0.660 \\
\bottomrule
\end{tabular}
\end{table}

Control error affects the three scales differently. On case118, the effect is
especially clear: bare control lowers AP from $0.673$ to $0.498$, whereas
protected control recovers $0.660$, close to the ideal-control result. The
case30 variation is comparatively small, while case14 shows substantial
sensitivity to introducing control imperfection but little separation between
the two modeled fidelity levels.

The important distinction from \Cref{app:hardware-noise} is therefore
architectural. QCS can benefit from using fewer measurement settings when
readout is noisy, but it has no analogous protection against errors in the
coherent operations that implement its learned sensing policy. Practical QCS
performance consequently depends on both the sensing interface and the
fidelity of the control layer.

\section{Quantum-Information Diagnostics}
\label{app:interpretability}

The scale results in \Cref{sec:results-rq3} raise a distinction that
classification accuracy alone cannot resolve: does QCS become less competitive
at larger scale because the quantum sensor state contains less useful
information, or because that information is not fully accessed by the chosen
readout and trained control model? We examine this question using
state-structure diagnostics and Fisher information.

\subsection{State coherence and purity}

As a basic state-level check, we compare the $\ell_1$-norm of off-diagonal
coherence and the purity $\mathrm{Tr}(\rho^2)$ between Normal and CTA states.
Both quantities change by less than $1\%$ relative across all three grid
scales. This is consistent with the physical regime considered here: the
topology change induces a comparatively small perturbation to the NV state,
so useful class information need not appear as a large change in a single
global state statistic.
These quantities are therefore calibration diagnostics rather than detection
metrics. 

\subsection{Controlled coherence perturbation}

To isolate information contained in coherence from information contained in
level populations, we apply the synthetic phase transformation
$
    U_\phi =
    \mathrm{diag}(1,e^{i\phi},e^{2i\phi})
$
to cached physical NV states. By construction, $U_\phi$ leaves all level
populations unchanged while modifying off-diagonal coherences. 
A population-only readout is correspondingly invariant to numerical precision,
with response $\leq1.1\times10^{-16}$ across all scales. In contrast,
Ramsey observables change by approximately $0.43$--$0.47$, while full
tomographic observables change by $0.70$--$0.78$. Thus, restricted
population measurement is structurally insensitive to state variations that
remain accessible to richer NV readout.
This diagnostic complements the single-setting comparison in
Appendix~\ref{app:reconstruction}: additional measurement structure can
increase information access even when all methods observe the same underlying
sensor state.

\section{Local-QCS state dimension}
Our primary architecture processes each NV qutrit independently before
classical fusion. Adding sensors therefore adds additional constant-size
$3\times3$ states rather than constructing a joint $K$-qutrit density
matrix. The state-storage requirement consequently scales as
$
    O(K)
$
for local processing, whereas exact storage of a fully joint $K$-qutrit
density matrix scales as
$
    O(9^K).
$
Thus, increasing the number of sensors does not increase the Hilbert-space
dimension of an individual trainable QCS module. This architectural
decoupling is important when interpreting the larger-grid experiments:
case118 is a larger sensing problem, but the local quantum state acted on by
each QCS module remains a single qutrit.

\subsection{Inference and memory scaling}

We benchmark a vectorized implementation in which the sensor dimension is
folded into the batch dimension while retaining the same shared QCS
parameters. The vectorized implementation was verified against the original
per-sensor implementation to numerical precision before benchmarking.

\begin{table}[h]
\centering
\caption{Inference and propagator-memory scaling with sensor count. Timing
uses 200 samples and is intended as a computational scaling benchmark rather
than an accuracy experiment.}
\label{tab:sensor-runtime}
\begin{tabular}{rccc}
\toprule
$K$ &
Inference time &
Per-sensor cost &
Propagator memory \\
\midrule
5    & 0.320 s & 320.3 $\mu$s & 0.6 MB \\
20   & 0.177 s & 44.2 $\mu$s  & 2.6 MB \\
100  & 0.514 s & 25.7 $\mu$s  & 13.0 MB \\
500  & 0.587 s & 5.9 $\mu$s   & 64.8 MB \\
1000 & 0.289 s & 1.4 $\mu$s   & 129.6 MB \\
2000 & 0.434 s & 1.1 $\mu$s   & 259.2 MB \\
\bottomrule
\end{tabular}
\end{table}

Memory grows approximately linearly with sensor count, reaching about
259\,MB at $K=2000$. Wall-clock timing is not monotonic because batching and
system utilization vary with $K$, but there is no evidence of the exponential
state-space growth that would occur under fully joint simulation.

\section{Detection accuracy versus sensor count}

Computational scalability does not imply accuracy is independent of $K$. We
therefore retrain Interleaved-QCS ($N=8$) from scratch on case118 CTA while
varying
$
    K\in\{5,10,20,50,100\}.
$

\begin{table}[h]
\centering
\caption{Interleaved-QCS accuracy versus sensor count on case118 CTA.
Each model is trained from scratch.}
\label{tab:accuracy-vs-k}
\begin{tabular}{rccc}
\toprule
$K$ &
AP (exact) &
AP (1000 shots) &
Train time \\
\midrule
5   & 0.696 & 0.528 & 988 s \\
10  & 0.731 & 0.697 & 2308 s \\
20  & 0.692 & 0.668 & 6670 s \\
50  & 0.705 & 0.672 & 14661 s \\
100 & \textbf{0.793} & \textbf{0.778} & 9070 s \\
\bottomrule
\end{tabular}
\end{table}

The dependence on $K$ is non-monotonic. Accuracy increases from $K=5$ to
$K=10$, decreases again at $K=20$, remains similar at $K=50$, and reaches
its highest value at $K=100$. Thus, the larger-scale regime is not difficult
simply because more sensors are being processed; adding sensors does not
systematically degrade the trained model.

Rather, case118 appears to define a different and more variable training
regime than case14 under the fixed training protocol. This distinction is
important for the interpretation of RQ3: \emph{grid scale, sensor count, and
trainability should not be treated as interchangeable quantities}. The
non-monotonic training times in \Cref{tab:accuracy-vs-k} are likewise not
interpreted as algorithmic scaling, since the runs were executed separately
under varying system load; \Cref{tab:sensor-runtime} provides the controlled
computational scaling measurement.

\section{Optimization Diagnostics at Larger Scale}
\label{app:realistic-training}

The Fisher-information analysis on native head
showed that the case118 sensor states retain information beyond that accessed
by conventional readout. We therefore test whether standard optimization
strategies can materially improve the amount of this information realized by
the trained Interleaved-QCS model with native head.
Our goal is diagnostic rather than exhaustive hyperparameter optimization.
Each method is evaluated with one reasonable configuration, allowing us to
test whether the larger-scale gap is immediately removed by commonly proposed
trainability interventions.


We evaluate five modifications to the baseline $N=10$ training procedure:
warm-start initialization from a converged case14 model, identity-oriented
initialization, progressive layerwise training, Quantum Natural Gradient
(QNG), and an auxiliary per-sensor loss.

\begin{table}[h]
\centering
\vspace{-0.4cm}
\caption{Optimization remedies for Interleaved-QCS ($N=10$) on case118
CTA.}
\label{tab:optimization-remedies}
\small
\begin{tabular}{lccc}
\toprule
Variant &
AP (exact) &
AP (1000 shots) &
Train time \\
\midrule
Baseline (random init., Adam)
& 0.733 & 0.686 & 2479 s \\
Warm start
& \textbf{0.739} & 0.693 & 2644 s \\
Identity initialization
& 0.736 & 0.677 & 2589 s \\
Layerwise $N:2\rightarrow10$
& 0.725 & 0.685 & 1637 s \\
Quantum Natural Gradient
& 0.736 & \textbf{0.701} & 2627 s \\
Auxiliary per-sensor loss
& 0.731 & 0.699 & 4010 s \\
\bottomrule
\end{tabular}
\end{table}

No individual intervention changes the larger-scale regime qualitatively.
Warm starting gives the highest exact-measurement AP, while QNG gives the
highest 1000-shot AP, but both improvements are modest. Identity
initialization and the auxiliary loss provide mixed changes, and progressive
layerwise training does not improve over the baseline.
These results do not identify a unique optimization mechanism, nor do they
establish a barren plateau: we do not perform the gradient-variance analysis
required for such a claim. They instead show that the observed scale gap is
not immediately removed by several standard optimization interventions.

\end{document}